\documentclass[aip, floatfix, reprint]{revtex4-1}
\usepackage{braket}
\usepackage{graphicx}
\usepackage{dcolumn}
\usepackage{bm}
\usepackage[dvipsnames]{xcolor}
\usepackage{xcolor}
\usepackage{amsmath}

\def\beg{\begin{gather}}
\def\eeg{\end{gather}}

\def\beq{\begin{equation}}
\def\eeq{\end{equation}}
\def\bea{\begin{eqnarray}}
\def\eea{\end{eqnarray}}
\def\ba{\begin{align}}
\def\ea{\end{align}}
\def\bs{\begin{split}}
\def\es{\end{split}}
\def\brcl{\begin{array}{rcl}}
\def\bccl{\begin{array}{ccl}}
\def\blcl{\begin{array}{lcl}}
\def\err{\end{array}}

\def\fatR{{\bf R}}

\def\dmc{{\mathrm{DMC}}}

\def\hf{{\mathrm{HF}}}
\def\xcb{{\mathrm{B3LYP}}}
\def\deltabq{{\Delta_{\xcb}^{\dmc}}}

\def\deltahq{{\Delta_{\hf}^{\dmc}}}

\def\rm{\mathrm}

\begin{document}

\title{Towards DMC accuracy across chemical space with scalable $\Delta$-QML}

\author{Bing Huang}
\email{bing.huang@univie.ac.at}
\affiliation{University of Vienna, Faculty of Physics, Kolingasse 14-16, 1090 Vienna, Austria}

\author{O. Anatole von Lilienfeld}
\email{anatole.vonlilienfeld@utoronto.ca}
\affiliation{Departments of Chemistry, Materials Science and Engineering, and Physics, University of Toronto, St. George Campus, Toronto, ON, Canada}
\affiliation{Vector Institute for Artificial Intelligence, Toronto, ON, M5S 1M1, Canada}
\affiliation{Machine Learning Group, Technische Universit\"at Berlin and Institute for the Foundations of Learning and Data, 10587 Berlin, Germany}
\author{Jaron T. Krogel}
\email{krogeljt@ornl.gov}
\affiliation{Materials Science and Technology Division, Oak Ridge National Laboratory, Oak Ridge, TN 37831, United States}
\author{Anouar Benali}%
\email{benali@anl.gov}
\affiliation{Computational Sciences Division, Argonne National Laboratory, Argonne, IL 60439, United States}

\begin{abstract}
{\color{black} 
In the past decade, quantum diffusion Monte Carlo (DMC) has been demonstrated to successfully predict the energetics and properties of a wide range of molecules and solids by numerically solving the electronic many-body Schr\"odinger equation. 
We show that when coupled with quantum machine learning (QML) based surrogate methods the computational burden can be alleviated such that QMC shows clear potential to undergird the formation of high quality descriptions across chemical space. 
We discuss three crucial approximations necessary to accomplish this: The fixed node approximation, universal and accurate references for chemical bond dissociation energies, and scalable minimal amons set based QML (AQML) models. 
Numerical evidence presented includes converged DMC results for over one thousand small organic molecules with up to 5 heavy atoms used as amons, and 50 medium sized organic molecules with 9 heavy atoms to validate the AQML predictions. 
Numerical evidence collected for $\Delta$-AQML models suggests 
that already modestly sized QMC training data sets of amons suffice to predict
total energies  with near chemical accuracy throughout
chemical space. 
}
\end{abstract}

\maketitle

\section{Introduction}

{\color{black} The predictive accuracy of quantum machine learning (QML) models trained on quantum chemistry data and used for the navigation of chemical compound space (CCS) is inherently limited by the predictive accuracy of the approximations used within the underlying quantum theory~\cite{huang2021abinitio}. Consequently, in order for QML models to achieve the coveted threshold of chemical accuracy ($\sim$ 1 kcal/mol average deviation of calculated from experimental measurements of atomization energies), it is necessary to rely on training data generated at least at the post-Hartree-Fock level, e.g.  CCSD(T)/CBS.
Unfortunately, the `gold-standard' in the field, CCSD(T)/CBS, generally imposes considerable computational cost due to steep prefactors and scaling $\propto O(N^7)$ ($N$ corresponding to system size)~\cite{MolecularElectronicStructureTheory}. 
As such, the routine generation of large high quality quantum data sets has remained elusive, even for relatively small organic molecules with only four or five `heavy' (second-row) atoms.
Here, we demonstrate for an exemplary sub-set of CCS (namely organic molecules) the usefulness of recently implemented and numerically more efficient Quantum Monte Carlo (QMC) methods for computing QML training data. 
Our numerical evidence indicates the possibility to routinely train QML models that achieve predictive power similar to QMC but at much reduced computational cost. }

QMC approaches solve the many-body electronic Schr\"odinger equation stochastically. QMC is general and applicable to a wide range of physical and chemical systems in any dimension or boundary condition etc. Amongst the most widely used flavors for electronic structure are the variational Monte Carlo (VMC)~\cite{bHammond1994,Lester1990} and diffusion Monte Carlo (DMC).\cite{foulkes01}. Both VMC and DMC are variational methods and allow to estimate the energy and properties of a given trial wavefunction without requiring to compute the matrix elements, posing no restriction on its functional form. Using the VMC algorithm, through stochastic numerical integration  scheme,  the  expectation  value  of  the  energy for  any  form  of  the  trial  wavefunction  can  be  estimated  by  averaging  the  local  energy over an ensemble of configurations distributed as $\psi^2$, sampled during a random walk in the configuration space using Metropolis~\cite{Metropolis1953} or Langevin algorithms~\cite{Reynolds1982}. The fluctuations of the local energy depend on the quality of the trial wavefunction, and they are zero if the exact wavefunction is used (zero-variance principle). DMC algorithm is very similar but the sampling goes beyond the $\psi^2$ distribution function by solving the Schrodinger equation in an imaginary time $\tau=it$ using a projector or a Green’s function based method. Any initial state $\ket{\psi}$, that is not orthogonal to the ground state $\ket{\phi_0}$ , will evolve to the ground state in the long time limit and any excited state will decay exponentially fast leading to the true ground state of the function. \\
\begin{equation}
\lim_{\tau \rightarrow \infty} \Psi(\textbf{R},\tau)=c_0 e^{-\epsilon_0\tau}\phi_0(\textbf{R})
\end{equation}
Where  $\ket{\psi}$ was expanded in eigenstates $\epsilon_i$ of the Hamiltonian as 
 \begin{equation}
 \ket{\psi}=\sum_{i=0}^{\infty}c_i\ket{\phi_0} , \hat{H}\ket{\phi_i}=\epsilon_i\ket{\phi_i}
  \end{equation}
  
 As mentioned, the sampling in DMC is not constrained to a specific distribution, takes into account all electronic correlations and therefore makes DMC a rigorously exact method. While this is true when solving for bosonic particles, solving for fermionic particles requires some approximations to remain computationally feasible and maintain the anti-symmetric nature of the wavefunction. Some of these approximations are controlled and can be rigorously extrapolated out (such as time steps, use of electron-core potentials or pseudopotential, etc...). The only uncontrolled source of error\cite{Anderson1980} is the fixed-node (FN) approximation introduced to suppress the fluctuations of the sign of the wavefunction (fermion sign problem). This approximation means that any proposed configuration changing the sign of the wavefunction is rejected, while any configuration lowering the local energy would be promoted. DMC being variational, if the positions of the nodes of the trial wavefunction are exact, the averaged local energy is rigorously the exact ground state energy. FN-DMC energies are an upper bound to the exact ground state energy\cite{Ceperley1980}. This implies that from a FN-DMC perspective, trial wavefunctions differ only by their nodal surface, and the best nodal surface leads to a lower energy and variance. \\
Despite the FN approximation, DMC was shown to reach successfully accuracy below the chemical accuracy threshold of 1 kcal/mol for chemical systems~\cite{Wagner2017,Dubecky2017,Burke2018,Dubecky2013,shulenburger2015} and a few tens of meV/unit-cells for solids within periodic boundary conditions~\cite{sorella2015,Foulkes2001,Shulenburger2013a,ArQMC_AnouarJCTC2014}.
Recently, multiple calculations using a selected Configuration Interaction (sCI) trial wavefunction have demonstrated how to systematically reduce the error from the fixed nodes\cite{Morales2012,caffarel2016,Scemama2018}. Nevertheless, these errors have proven to be significant only for open shell or multi-reference molecules.\cite{Burke2018,Wagner2013}  \\

{\color{black}From above discussion, we see that on the one hand, stochastic numerical sampling permits  independent evaluationsy, making the method embarrassingly parallel and highly efficient for high performance computing (HPC).
On the other hand, accuracy is a direct consequence of the quality of the fixed nodes in the trial wavefunction: If the nodes are exact, the method is rigorously exact.
Using DMC energies as reference for QML models will then boost efficiency by several orders of magnitude,
since the property of any new out-of-sample query compound can be predicted after training, solely based on inference from the DMC information stored in the training data.
In order to retain the predictive accuracy of reference data, however, a significant amount of training data can be necessary.
This issue is essentially caused by the use of random selection of training instances which should be representative of query compounds. 
To rise to this challenge, some of us (BH, OAvL) recently introduced the amon (A) based QML method which enables a dramatic reduction in training set size as well as size of training molecules~\cite{Amons}. 
 }


{\color{black}Amons correspond to systematically fragmented entities of query target molecules, containing an increasing number of heavy atoms (typically no more than 7). They can be seen as effective building blocks of target compounds with atomic states being perturbed according to their chemical environment.} With amons used as training set, QML models trained on the fly (AQML) represent a scalable approach which can be applied throughout chemical space to predict quantum properties of large molecules.

For this study, we combine DMC reference calculations with AQML and $\Delta$-AQML models,
and we numerically demonstrate the feasibility of these approaches to achieve chemical accuracy, $\Delta$-AQML in particular,
by making use of a dictionary of 1175 small amons of QM9~\cite{QM9} with up to \emph{only} 5 heavy atoms (not counting hydrogens),
together with reference energies calculated at much cheaper levels of theory,
including mean-field theory (Hartree-Fock (HF) and density functional theory (DFT) level using various levels of approximations according to Jacob's ladder)
and M{\o}ller–Plesset perturbation theory (MP) to second (MP2).


\section{Data-sets} \label{sec:db}

1175 unique \emph{amon graphs} (i.e., molecular graphs) with up to 5 atoms were firstly identified by application of the amon-selection algorithm~\cite{Amons} (also briefly summarized below in the methodology section~\ref{sec:meth_amons}) to QM9~\cite{QM9,ReymondChemicalUniverse3} molecules, with SMILES strings as the only input. Then for each amon one hundred conformers were sampled using RDKit~\cite{rdkit}, optimized by MMFF94 force field and only the global minimum configuration (i.e., lowest force field energy conformer) was chosen. The geometries of the thus-selected \emph{amon conformers} were further optimized at the level of theory B3LYP/Def2TZVP. 
Based on these geometries, single point energies were calculated at multiple levels of theory, including HF, PBE, PBE0, B3LYP, MP2 (with basis cc-pVTZ) and DMC (see the next section for details).
The resulting dataset can be seen as a compact dictionary of small molecules, which is to be looked up later in AQML for any query molecule of larger size.

For test purpose of QML models, 50 molecules all made up of 9 heavy atoms are randomly drawn from the QM9 dataset.
Geometries were optimized at the same level of theory as for amons,
followed by single point energy calculations by all levels of theory mentioned above, including DMC.
For a depiction of all test molecules, see Fig.~\ref{fig:test50}.


\section{Computational details}
We used B3LYP/Def2TZVP for geometry optimization as implemented in the Gaussian 09~\cite{g09} code.
For HF, DFT and post-HF (MP2) single point energies, we switched to the cc-pVTZ basis,
and used instead Molpro2018~\cite{molpro} with cc-pVQZ-jkfit density-fitting basis to speed-up computations of both Coulomb and exchange integrals.\\
For DMC calculations, we used a trial wavefunction with a Slater Jastrow form~\cite{Schmidt1990}
 \begin{equation}
\Psi_T(\vec{R}) = \exp\left[\sum_i J_i(\vec{R})\right]\sum_k^M C_kD_k^{\uparrow}(\varphi)D_k^{\downarrow}(\varphi)
\end{equation}
Where $D_k^{\downarrow}(\varphi)$ is a slater determinant expressed in terms of single particle orbitals (SPO) $\varphi_i=\sum^{N_b}_l C_l ^i \Phi_l$ . We use DMC to evaluate the total and formation energies of all molecules in data-sets, as implemented in the QMCPACK code~\cite{kim2018qmcpack,kent2020qmcpack}. Our trial many-body wave functions are constructed with the product of the Jastrow functions and a single Slater determinant from Hartree Fock, PBE\cite{PBE} PBE0\cite{PBE,PBE0} and B3LYP\cite{B3LYP1,B3LYP2,B3LYP3,B3LYP4} Kohn-Sham (KS) orbitals obtained from the PYSCF\cite{pyscf} package. 
Using a variant of the linear method of Umrigar and co-workers~\cite{umrigar07}, up to 40 variational parameters including one-body, two-body and three-body Jastrow factors are optimized within VMC, after which we can obtain the ground-state energies with diffusion Monte Carlo (DMC) under the fixed node approximation~\cite{foulkes01}. For all molecules, we used nodal surfaces coming from HF and aforementioned DFT functionals as a way of assessing the quality of the trial wavefunction. Jastrow parameters were optmized independently for each molecule and each trial wavefunction. All calculations used a 0.001 time-step holding an error within error bars of the extrapolated 0 time-step. This was verified by randomly selecting 10 molecules of different size and running the time-step extrapolation. With such small time-steps, we increased the size of decorrelation time to avoid auto-correlation. Each molecule used 4096 walkers and 2000 blocks to insure convergence. Using the resources of the ALCF-Theta supercomputer (Cray XC40, with Intel Xeon Phi KNL processors), each molecules was run on 32 nodes and 128 threads (2 hyperthreads ) for 1 hour of compute time or a total of 9.6M core-hours.\\

For all QML models, we rely on a local representation called
atomic Spectrum of London and Axilrod-Teller-Muto potential (aSLATM)~\cite{Amons},
with a weight of 1 and 1/3 for the 2-body London potential and 3-body Axilrod-Teller-Muto potential respectively (Note that 1-body terms are not necessary),
to describe atoms in molecules (i.e., atomic environments).
Default 1D grids were used, as was implemented in the original \texttt{aqml} code (available at https://github.com/binghuang2018/aqml),
with grid spacing 0.05 \r{A} (rad) for the 2-body (3-body) potential,
ranging from 0.2 to a maximal atomic cutoff of 4.8 \r{A} for the 2-body part
and 0 to $\pi$ for the 3-body part, respectively.
A smearing width of 0.05 \r{A} (rad) was used for the normalized 1D Gaussian distribution centered on each bond distance (bond angle) within the atomic cutoff.
$L_2$ norm was used to compute the distances between two atomic environments and Gaussian kernel was used to measure their similarity,
with atom type (characterized by nuclear charge) dependent kernel width set to the maximal value of aSLATM distance between all pairs of atomic environments (of the same kind) divided by $\sqrt{2\ln 2}$.
A universal regularization parameter of $10^{-4}$ was used to reduce the complexity of all QML models.

\section{Methodology}

\subsection{From amons to $\Delta$-AQML} \label{sec:meth_amons}

To provide sufficient context, we now briefly summarize the key ideas underlying amons and their use within $\Delta$-AQML models. The interested reader is referred to the original papers~\cite{Amons, DeltaPaper2015} for further details.

The amons approach attempts to mitigate the curse of dimensionality in CCS
through selection of the smallest possible, yet ``optimal'', training set on-the-fly after having been provided a given specific query test molecule (or a set). 
The amons selection procedure~\cite{Amons} can be roughly divided into three major steps:
a) Perceive the connectivity graph $G$ of any query based on its 3D geometry.
b) Collect all isomorphic subgraphs \{$G_i$\} of $G$ with no more than $N_I$ heavy atoms ($N_I$ is set to 5 in this study) based on an efficient tree enumeration algorithm.
Subgraph isomorphism help retain all hybridization states of atoms during fragmentation (hydrogens are added to heavy atoms when necessary).
c) Perform geometry relaxation for each fragment with some force field and subsequently quantum chemical approach,
with dihedral angles constrained to match that in the query molecule so as to avoid too much change in conformational degrees of freedom.
The resulting fragments, if unique and survived (i.e., no dissociation or graph change after geometry relaxation), are then selected for the amon database.

In case of neglect of amon conformers, i.e., precise geometry information of the query is disregarded by providing only the molecular graph of the query (which allows further reduction in training set size)
the last step (step c) is replaced by global minima search based on force field methods,
followed by geometry relaxation at some quantum chemical level of theory.
Given the complete set of amons, any query molecule,
even if even substantially larger than amons in molecular size,
can then be predicted to high accuracy by QML models,
when used in conjunction with atomic representations. 
Note that the overall number of amons for any given query, can be rather small, in
particularly for highly regular systems  exhibiting repeating patterns or periodicity, e.g. polymers, peptides, or crystals.

Nonlinear kernel based ridge regression (KRR) has been shown to be a rather robust regressor in previous studies~\cite{Amons}.
Within KRR, the energy of a query $q$ is a sum over weighted kernels,
\beq 
E^{\mathrm{est}}_q =\sum_{i=1}^{N} \alpha_{i} K\left(\mathbf{M}_q, \mathbf{M}_{i}\right) =\sum_{i=1}^{N} \alpha_{i} K\left(q, i\right)
\eeq 
where $\mathbf{M}_q$ is the molecular representation of $q$ and the molecular kernel $K(q,i)$ measures the similarity between $q$ and the $i$-th training molecule (the overall set size is $N$).
Regression coefficients $\{\alpha_i\}$ are to be obtained from training.

To achieve scalability, i.e., generalization to larger molecules after training on small ones, the molecular kernel is expressed as summations of atomic kernels, i.e.,
\begin{gather} 
K\left(\mathbf{M}_q, \mathbf{M}_{i}\right) = \sum_{I\in i} \sum_{Q\in q} k\left(\mathbf{M}_q^Q, \mathbf{M}_{i}^I\right) \\ 
k\left(\mathbf{M}_q^Q, \mathbf{M}_{i}^I\right) = \delta_{Q,I}  \exp \left[-\frac{1}{2 \sigma^{2}} || \mathbf{M}_q^Q -  \mathbf{M}_{i}^I ||_2^2 \right]
\end{gather}
where $\mathbf{M}_q^Q$ denotes the aSLATM representation of atom $Q$ in molecule $q$,
$\sigma$ is the kernel width,
$||\cdot||_2$ is the $L_2$ norm (i.e., Euclidean distance),
$\delta_{Q,I}$ is the Kronecker delta
\beq 
\delta_{Q,I} = \begin{cases}
1, & Z_I=Z_J \\
0, & Z_I\ne Z_J
\end{cases}
\eeq 
with $Z_Q$ being the nuclear charge of atom $Q$.

To determine $\left\{\alpha_{i}\right\}$, we solve the following minimization problem,
\beq 
\min _{\boldsymbol{\alpha}} \sum_{i}\left(E^{\mathrm{est}}\left(\mathbf{M}_{i}\right)-E_{i}^{\mathrm{ref}}\right)^{2}+\lambda \sum_{i} \alpha_{i}^{2}
\eeq
where the second part corresponds to a regularization term with coefficient $\lambda$,
limiting the norm of regression coefficients and thereby controlling the model complexity.

For given kernel width $\sigma$ and regularization parameter $\lambda$, the explicit solution to the minimization problem
is given by 
\beq 
\boldsymbol{\alpha}=(\mathbf{K}+\lambda \mathbf{I})^{-1} \mathbf{E}^{\mathrm{ref}}
\eeq 
where $\mathbf{K}$ is the kernel matrix of all training molecules, $\mathbf{I}$ denotes the identity matrix,
$\mathbf{E}^{\mathrm{ref}}$ represents the $N$ by 1 matrix with the $i$-th entry being the reference energy of the $i$-th training molecule.

Usually, one does not directly use reference energies to form $\mathbf{E}^{\mathrm{ref}}$,
instead, shifting to some common grounds of atomic energies is necessary,
so as to centralise the data, 
making the magnitudes of the reference energies considerably smaller and easier to learn.
The common ground corresponds to the so-called ``dressed-atom'' (DA) and its energy could be obtained through least square regression of a simple linear model, i.e.,
\beq 
E^{\mathrm{ref}}_i =  \sum_A n^i_A \varepsilon_A^{\mathrm{ref}}
\eeq 
where $n_A^i$ is the number of atoms of type A in the $i$-th traning molecule and $\varepsilon_A$ corresponds to the dressed-atom energy of A.
This null model is termed ``dressed-atom'' model hereafter,
and is vital for treating total energy or relative energies (e.g., atomization energy, free energy of formation, etc.) on an equal footing.
To summarize, $E^{\mathrm{ref}}_i - \sum_A n^i_A \varepsilon_A^{\mathrm{ref}}$ is always used for training and test for (single-level) AQML models,
and the total energy can be easily recovered by adding $\sum_A n^i_A \varepsilon_A^{\mathrm{ref}}$.

The amons-based QML framework (AQML) described above generally allows for very effective extrapolation to larger test molecules
after training on amons (including conformers) made up of 7 heavy atoms (i.e., $N_I=7$) at most.
Throughout this research, we have amons without any conformer and $N_I$ is limited to 5 at most,
ending up with way fewer amons than the $N_I\leq 7$ case
and faster training,
at a price of, to some extend, compromised extrapolation capability.
Targeting chemical accuracy with such small training set, the $\Delta$-AQML model may come to rescue,
which combines the idea of amons and $\Delta$-ML~\cite{DeltaPaper2015}.

Within $\Delta$-AQML, the energy delta between two reference levels of theory is used for training and test.
For instance, for a $\Delta$ model involving two reference levels B3LYP and DMC,
$\Delta E^{\mathrm{DMC-B3LYP}}=E^{\mathrm{DMC}} - E^{\rm{B3LYP}}$ is used.
Accordingly, the null or ``dressed-atom'' (DA) model becomes
\bea \label{eq:lsq1}
    \Delta E^{\mathrm{DMC-B3LYP}}_i &=&  \sum_A n_A^i \varepsilon_A^{\rm{DMC-B3LYP}} 
\eea
And after regression of coefficients $\{ \alpha_i \}$, the DMC energy for any query molecule $q$ can be estimated by (unless otherwise stated, all DMC energies used in QML models are based on B3LYP nodal surface):
\begin{align}
E^{ \rm{est,DMC} }_q = &E^{\rm{B3LYP}}_q + \Delta E_q^{\mathrm{est,DMC-B3LYP}}
\end{align} 
Similar to the AQML case, one needs to add $\sum_A n_A^q \varepsilon_A^{\rm{DMC-B3LYP}}$ to obtain the estimated DMC total energy.

Hereafter, we use QML as a general name for any ML models. We use AQML to denote any single level QML model using amons as training set, $\Delta$-AQML for QML models taking as input reference energies of amons calculated by two levels of theory,
one cheap (coarse) and the other expensive (high in accuracy).
An AQML model using reference data calculated at, say HF, is termed HF trained AQML.
As for $\Delta$-AQML models, model name is simply the concatenation of levels of theory involved seperated by a minus sign and appended by AQML,
with the first (second) level of theory as baseline (target),
e.g., for the $\Delta$-AQML model as shown above through the equations,
the model is named $\deltabq$-AQML, or $\deltabq$ for short.

To assess the average performance of QML models for multiple molecules, mean absolute error (MAE, calculated as $ \sum_{q=1}^{N_{\rm{test}}} |E^{\rm{est}}_q-E^{\rm{ref}}_q |/N_{\rm{test}}$), root mean squared error (RMSE, i.e., $\sqrt{  \sum_{q=1}^{N_{\rm{test}}} (E^{\rm{est}}_q -E^{\rm{ref}}_q)^2/N_{\rm{test}} }$) and maximal absolute error (MaxE, that is, $\max_q |E^{\rm{est}}_q -E^{\rm{ref}}_q | $) are used.
When the performance of QML models for individual molecule is concerned, the signed error, which is equal to $E^{\rm{est}}_q-E^{\rm{ref}}_q$, is used instead.

\subsection{Training set selection} 
Different to the setup in the original amons paper~\cite{Amons}, here we have a prepared amons dictionary (see section~\ref{sec:db} for details),
and for any new query we have to look up the dictionary to find the ones that best match local geometries in the target.

A naive look-up strategy is to consider simply the molecular graph (or bond connectivity matrix).
More specifically, the subgraph isomorphism could be used as the only criteria for amons selection,
as in the amons generation procedure taking molecular graph as input, as described in the methodology section.
This naive approach is expected to work for very rigid molecules,
for which, minor difference exists between the full amons set and the graph amons set (that is, very few or no amon conformers),
e.g., conjugated alkenes.

For more general cases, where conformational degrees of freedom comes into play,
the simple graph approach is not enough
and local geometries must be concerned.
Aiming for speed and robustness, we use the inverse distance matrix $\mathbf{R}^{-1}$,
with 1.0 as the diagonal and the $P,Q$-th entry being the inverse of the interatomic distance between atom $P$ and $Q$, i.e., $1/R_{PQ}$.
The matrix infinity norm (same as 1-norm for the symmetric $\fatR^{-1}$ matrix)
is considered as the measure of the distance between the $i$-th amon in the dictionary and any matched local fragment in the query $q$ (labeled as $q_j$. Note that it is possible to have multiple matches), i.e.,
\bea
d(i,q_j) &=& \|\mathbf{R}^{-1}_i - \mathbf{R}^{-1}_{q_j}\|_{\infty}\\
&=& \max_{1 \leq Q \leq n_{q_j}} \sum_{P=1}^{n_{q_j}} |(\fatR^{-1}_i)_{PQ} - (\fatR^{-1}_{q_j})_{PQ}|
\eea 
where $P,Q$ run over all matched atoms and $n_{q_j}$ the total number of atoms in $q_j$ (Note that $n_{q_j}<n_i$).
The final distance between the $i$-th amon and the target $q$ is calculated as
\beq 
d_{iq} = d(i,q) = \min_{j} d(i,q_j) 
\eeq 
If the $d_{iq}$ is below a threshold of 0.5, the $i$-th amon is selected for training; otherwise, skipped.

\section{Results and Discussion}
\subsection{Assessing the accuracy of total and dressed atomic DMC energies}

As described previously, DMC is variational and its energy remains an upper bound to the exact energy of the system. Better nodal surfaces, given by better trial wavefunctions, always lead to lower total energies while an exact nodal surface would lead to the exact ground state energy.  The quality of the nodal surface therefore controls the accuracy of a DMC calculation. A direct optimization of the position of the nodes is unfortunately impossible as the number of optimizable parameters grows exponentially with the number of electrons in the system. More recently,  promising results were obtained with trial-wavefunction-free variational Monte Carlo (VMC)\cite{pfau2019abinitio} using Fermionic neural networks, however, such methods do not apply (yet) to large systems or to DMC. Nevertheless, it is still possible to improve the quality of a nodal surface by increasing its complexity, e.g. by using a multideterminant trial wavefunction\cite{Caffarel2016,Morales2012}), by adjusting the fraction of exact exchange with a DFT hybrid functional to minimize the DMC energy\cite{Busemeyer2016}. In this study, we used 4 different trial wavefunctions for all molecules in the data set. To the best of our knowledge, this is the most exhaustive and systematic use of various nodal surfaces applied to a large set of molecules to date.

\begin{figure}[t!]
\includegraphics[width=\columnwidth]{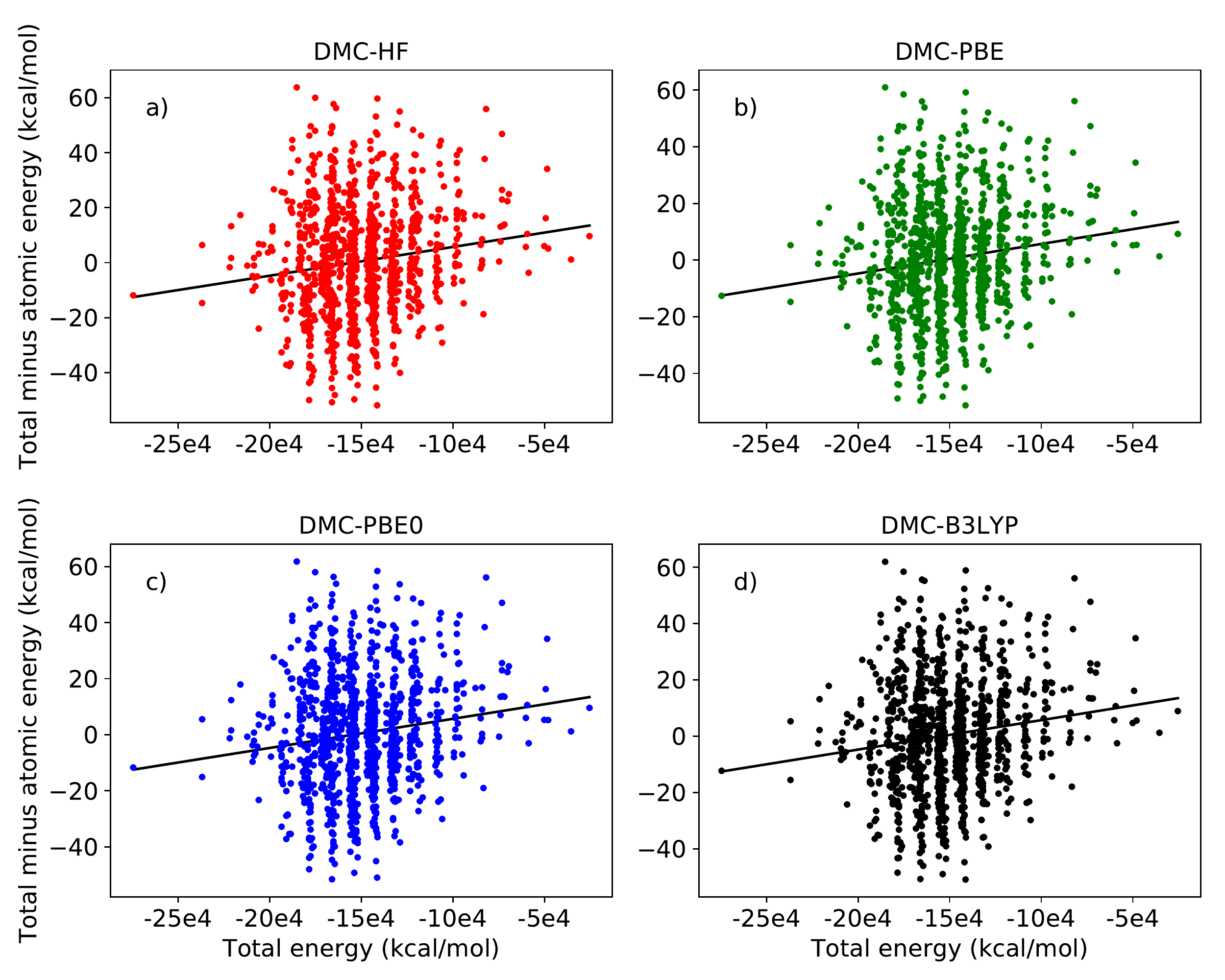}
\caption{Total DMC molecular energies for the $\{N_I\le 5\}$ set with the self-consistent dressed atomic energies removed, plotted vs. the total energies.  Results are shown for (a) DMC-HF, (b) DMC-PBE, (c) DMC-PBE0, and (d) DMC-B3LYP.  DMC shows detailed agreement in the residual energies regardless of the source of nodes.  Linear least squares fits are shown in black.}
\label{fig:dmc_tma}
\end{figure}

Diffusion Monte Carlo gives a highly consistent representation of the compositional energy landscape spanned by the space of training molecules.  This can be seen directly when considering the beyond atomic contributions to the energy that are present in DMC.  Figure \ref{fig:dmc_tma} contains DMC total energies calculated with HF, PBE, PBE0, and B3LYP nodes for the entire $\{N_I\le 5\}$ set of molecules with dressed atomic energies--calculated self-consistently within each set--removed.  First, it is immediately apparent that the beyond atomic contributions to the total energy agree in a detailed way across the four inputted nodal structures.  The pattern expressed in each panel of figure \ref{fig:dmc_tma} is essentially within set target for the machine learning models just described.  The detailed agreement between the patterns demonstrates that any of the DMC reference sets could be used effectively to model the beyond atomic variations in the compositional energy landscape.  This also indirectly highlights the crucial role the derived atomic energies play in our model representations.  The DMC total energies span a space ranging from roughly $-2.5\times 10^5$ to $-5\times 10^4$ kcal/mol.  By removing the dressed atomic energies, the residual energies span a space ranging from about $-40$ to $60$ kcal/mol, a reduction of more than three orders of magnitude.  This means that the dressed atomic energies capture the vast majority of the variation within the compositional energy landscape, and, as we will show below, they also capture the bulk of the error present in the energy landscape of a particular theory.  This property is essential to propagate the accuracy present in DMC to the more affordable QML models.



The DMC variational principle can be used to directly select the most accurate energies corresponding to alternative sets of trial wavefunctions.  By taking an average over the $\{N_I\le 5\}$ set of molecules, 
we find the lowest single reference DMC energies to be provided by B3LYP nodes, followed closely by PBE0. Single reference trial wavefunctions based on PBE0, PBE, and HF produce energies that are higher than those of DMC-B3LYP by 0.08(2), 0.26(2), and 4.74(2) kcal/mol on average. 
We can further assess the accuracy of the nodal surface, by increasing the complexity of the trial wavefunction. Such assessment is beyond the scope of this work, but is being investigated by us and will be published independently.

\begin{figure}[t!]
\includegraphics[width=\columnwidth]{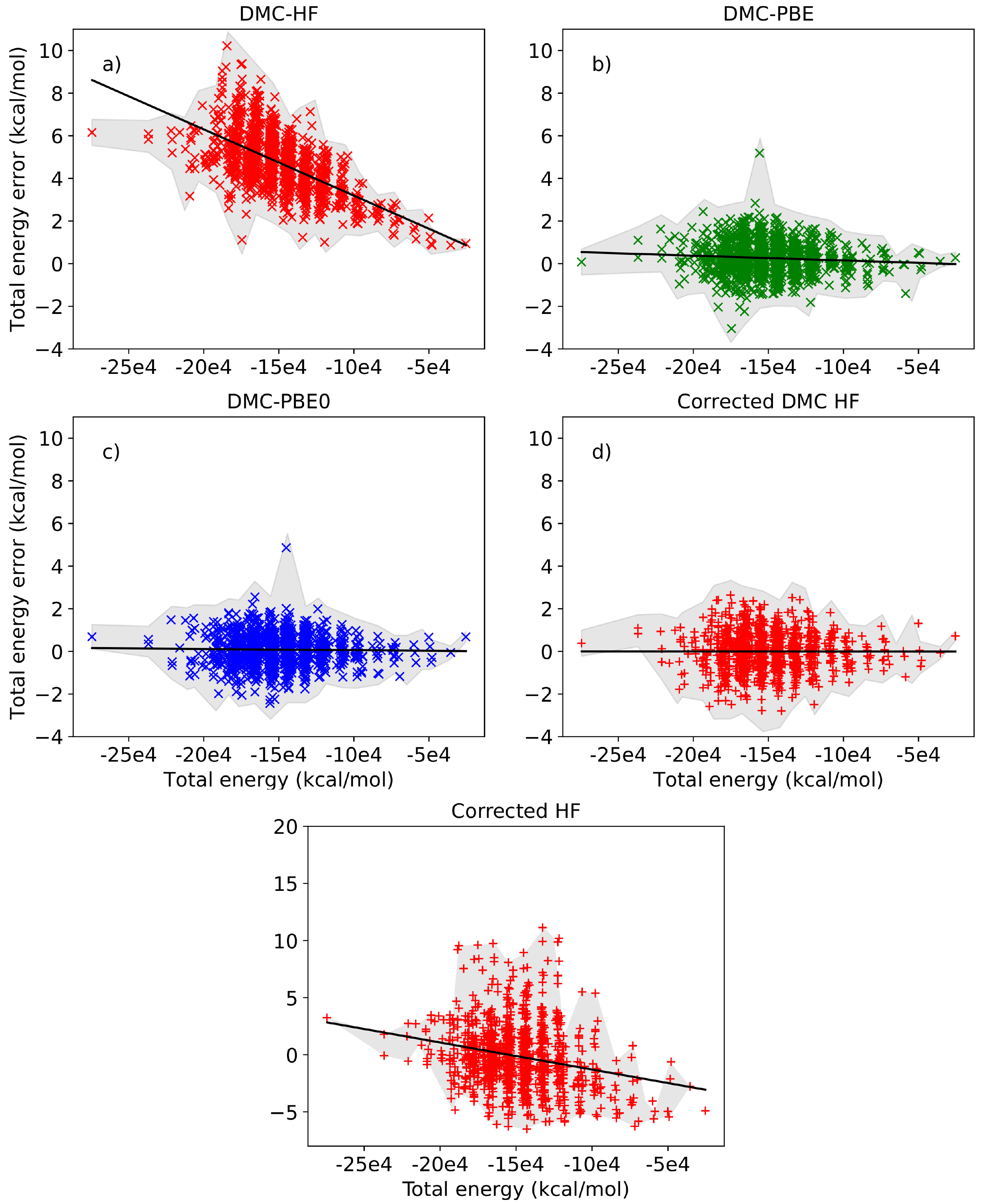}
\caption{Errors in DMC total energies as measured against the DMC-B3LYP reference for the $\{N_I\le 5\}$ set.  Results are shown for (a) DMC-HF, (b) DMC-PBE, and (c) DMC-PBE0.  In (d) and (e), corrected energies are obtained by adding the difference between the DMC-B3LYP and DMC-HF dressed atomic energies to the raw DMC-HF total energies (d) and the raw HF total energies (e).  In all cases, the shaded region indicates the 1 $\sigma$ statistical uncertainties of the outlying energies.  Linear least squares fits are shown in black.}
\label{fig:dmc_terr}
\end{figure}
 
 Before considering results for the models themselves, we can gain further insight into the correction mechanisms afforded by the models by inspecting the relative accuracy of the DMC single reference energies, and the degree of increase in accuracy afforded by employing a simple correction based on the dressed atomic energies.  Figure \ref{fig:dmc_terr} panels (a)-(c) show the error in DMC-HF, DMC-PBE, and DMC-PBE0 energies  measured relative to the DMC-B3LYP reference as a function of the molecular total energy.  The DMC-PBE and DMC-PBE0 results are essentially unbiased as a function of total energy, as shown by the nearly horizontal linear least squares fits shown in black.  The energies arising from these nodal surfaces also agree closely with DMC-B3LYP across all molecules with a RMSE of 0.92(1) kcal/mol across the full set.  The DMC-HF energies, however, clearly show a bias for nearly all molecules compared to DMC-B3LYP.  This bias grows significantly as the total energy of the molecules increases.  Training a QML model on the DMC-HF energies would therefore bias the model to give increasingly incorrect predictions for these molecules.  The bulk of the error presented by DMC-HF, as well as by simpler density functional theories, is captured by the dressed atomic energies that are derived self-consistently within each theory.  We demonstrate this in figure \ref{fig:dmc_terr}(d), where we apply the simplest possible correction based on the atomic energies.  The corrected results shown there are found simply by adding the difference in DMC-PBE0 and DMC-HF atomic energies from the DMC-HF total energies.  A correction of this type, which involves only the five dressed atomic energies, remarkably removes nearly all of the error present in DMC-HF and produces instead a set of unbiased molecular energies with accuracy similar to DMC-PBE0 or DMC-B3LYP. Similar correction can be applied to the raw HF total energies and shows in figure \ref{fig:dmc_terr}(e) a significant improvement of the HF error despite the simplistic model used. 
 
\begin{figure}[t!]
\includegraphics[width=\columnwidth]{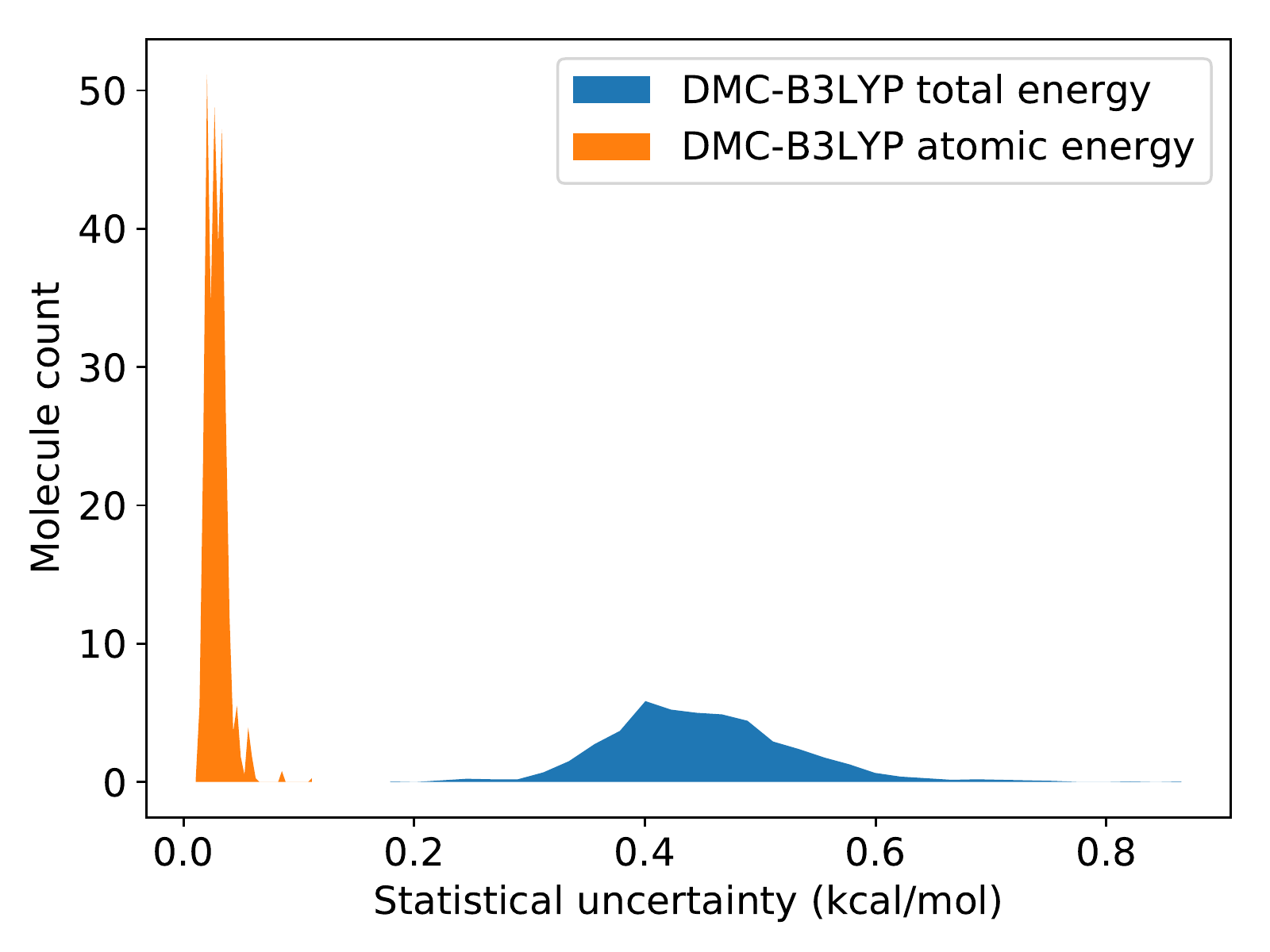}
\caption{Distribution of the 1 $\sigma$ statistical uncertainties of the DMC-B3LYP molecular total energies (blue) and molecular energies reconstructed from the DMC-B3LYP dressed atomic energies (orange).  The fitting/learning process greatly reduces the statistical uncertainty of the output relative to the input.}
\label{fig:dmc_tunc}
\end{figure}
 
 A final benefit to using dressed atomic DMC energies as the basis for QML models is decreased statistical uncertainty.  In figure \ref{fig:dmc_tunc} we show the distribution of DMC-B3LYP statistical uncertainties for the $\{N_I\le 5\}$ set of molecules (blue).  The distribution is peaked near a median statistical uncertainty of 0.45 kcal/mol.  When reconstructing the set of molecular energies from the dressed atomic energies, the statistical uncertainty for each molecule is greatly reduced (shown in orange).  The median of this distribution, which directly informs the QML models, is instead 0.03 kcal/mol, which is a substantial increase in precision.  To the extent that the QML models compress information during the learning process from DMC data, the predicted outputs from the models will enjoy an analogous reduction of statistical uncertainty beyond the resolution of the inputted DMC datasets.

\begin{table} \label{tab:errors}
\centering
\caption{Direct and $\Delta$-AQML models prediction errors [mean absolute error (MAE), root mean squared error (RMSE), maximal absolute error (MaxE)] for total energies (in kcal/mol) of 50 test QM9 molecules,
trained on the respectively largest amons set possible (on average 21 amons).
In each row, bold numbers denote lowest errors. } 
\begin{tabular}{llllll} 
\hline\hline
   & HF                    & PBE                  & B3LYP                      & MP2                     & DMC              \\ 
\hline
MAE    & 6.42  & 5.94  & 6.06   & \textbf{5.31}  &  5.78    \\
RMSE   & 9.92  & 8.81  & 9.11   & \textbf{8.14}  &  8.67    \\
MaxE   &42.60  &37.59  &39.51   & \textbf{35.75} &  36.32   \\
                               &                       &                      &                        &                                              &                \\
 & $\Delta _{\rm{HF}}^{\rm{DMC}}$                  & $\Delta _{\rm{PBE}}^{\rm{DMC}}$                & $\Delta _{\rm{B3LYP}}^{\rm{DMC}}$                    & $\Delta _{\rm{MP2}}^{\rm{DMC}}$                         &                              \\
\cline{1-5}
MAE  & 1.99  & 1.89  & \textbf{1.56}  & 1.70           &               \\
RMSE & 2.80  & 2.56  & \textbf{2.02}  & 2.05           &               \\
MaxE & 11.00 & 7.24  & 6.39           & \textbf{5.90}  &               \\
\hline
\end{tabular}
\end{table}

\subsection{Amon based QML and $\Delta$-AQML}

\subsubsection{Mean error analysis}\label{sec:errana}
Mean errors, as well as MaxE for all QML models are summarised in Table I.
To understand the results of $\Delta$-AQML models, it is natural to start from AQML ones. 

Once trained on the respective largest amons set possible for each of the test molecules,
all AQML models yield a MAE of $\sim$6 kcal/mol for the 50 test molecules.
This mean error is roughly of hybrid DFT quality,
is already impressive, in our opinion, considering that only 21 amons on average are used for training,
each made up to 5 heavy atoms at most (i.e., $N_I\leq 5$).
Though this error is still somewhat ``distant'' from the highly coveted chemical accuracy ($\sim$ 1 kcal/mol),
it is expected to converge to a much lower MAE when larger amons of size up to $N_I=7$ are added for training,
as has been reported for 11k QM9 molecules with a MAE of $\sim$1.6 kcal/mol~\cite{Amons}.

Meanwhile, it is intriguing to compare the MAE's from different AQML models:
HF energies turns out to be more difficult to learn than any other post-HF based AQML models
(i.e., relative performance: MP2 trained $>$ HF trained AQML),
where the corresponding reference level of theory can account for correlation energy to some extent.
This suggests that the energy of atom in molecule from correlated post-HF model is more transferable than its HF counterpart across different molecules,
as the performance of QML models based on atomic representation relies on the validity of locality assumption of atom in molecule,
and correlated model typically promotes electrons from the highly delocalized canonical molecular orbitals (MO) to virtual MO's,
which pushes electrons closer (back) to the nucleus (as in the atomic orbital) and
results in a more localized picture of electrons (as well as the atom on which these electrons are ``centered'') in molecule.
The finding that inclusion of virtual MO's improves MO localization for highly entangled cases (e.g., benzene molecule),
as in the partly occupied Wannier function~\cite{thygesen2005partly},
also help support the reasoning above. 

Similarly, DFT trained AQML models exhibit reduced MAEs than the HF trained AQML model,
since DFT can also explicitly account for electron correlation. 
A possible reason for this observation could be that the local or semi-local nature of the exchange-correlation potential serves as a driving force that renders the
electronic system more localized than in the HF case.
In spite of B3LYP corresponding to a higher rung of Jacob's Ladder,
the MAE of PBE vs.~B3LYP trained AQML is almost identical.
Direct comparison of DFT trained AQML and AQML models trained on post-HF methods, however, is not quite meaningful, as the content of electron correlation in different DFT methods is hard to trace (subtraction of dressed-atom energies further complicates the analysis),
due to their empirical nature (i.e., the exact functional form of $E_{xc}[\rho]$ is unknown and has to be approximated, with some parameters fitted to experimental or computational data).
DMC trained AQML produces a MAE of 5.78 kcal/mol which is in between that of DFT trained AQML and MP2 trained AQML.

While sticking to the same set of amons (with $N_I\leq 5$), the MAE's drop by typically more than two thirds when shifting from DMC trained AQML to $\Delta$-AQML models,
using the energy of DMC as target level together with a much cheaper level of thoery as baseline.
For instance, DMC trained AQML yields a MAE of 5.78 kcal/mol, which decreases to 1.99 kcal/mol using HF energy as baseline
in the $\deltahq$-AQML model.
Among the several $\Delta$-AQML models (see table 1), the $\deltabq$-AQML model offers the smallest MAE, i.e., 1.56 kcal/mol.



RMSE's in general follow similar trends as MAE's ($\sim$1.5 times of MAE in magnitude), while MaxE's (maximal absolute error) are considerably larger,
indicating lack of local atomic environments in amons of the outliers.
We will come back to this point later through a detailed signed error analysis for each test molecule.

\subsubsection{Learning curves}

\begin{figure*}
\centering
\includegraphics[scale=1.0]{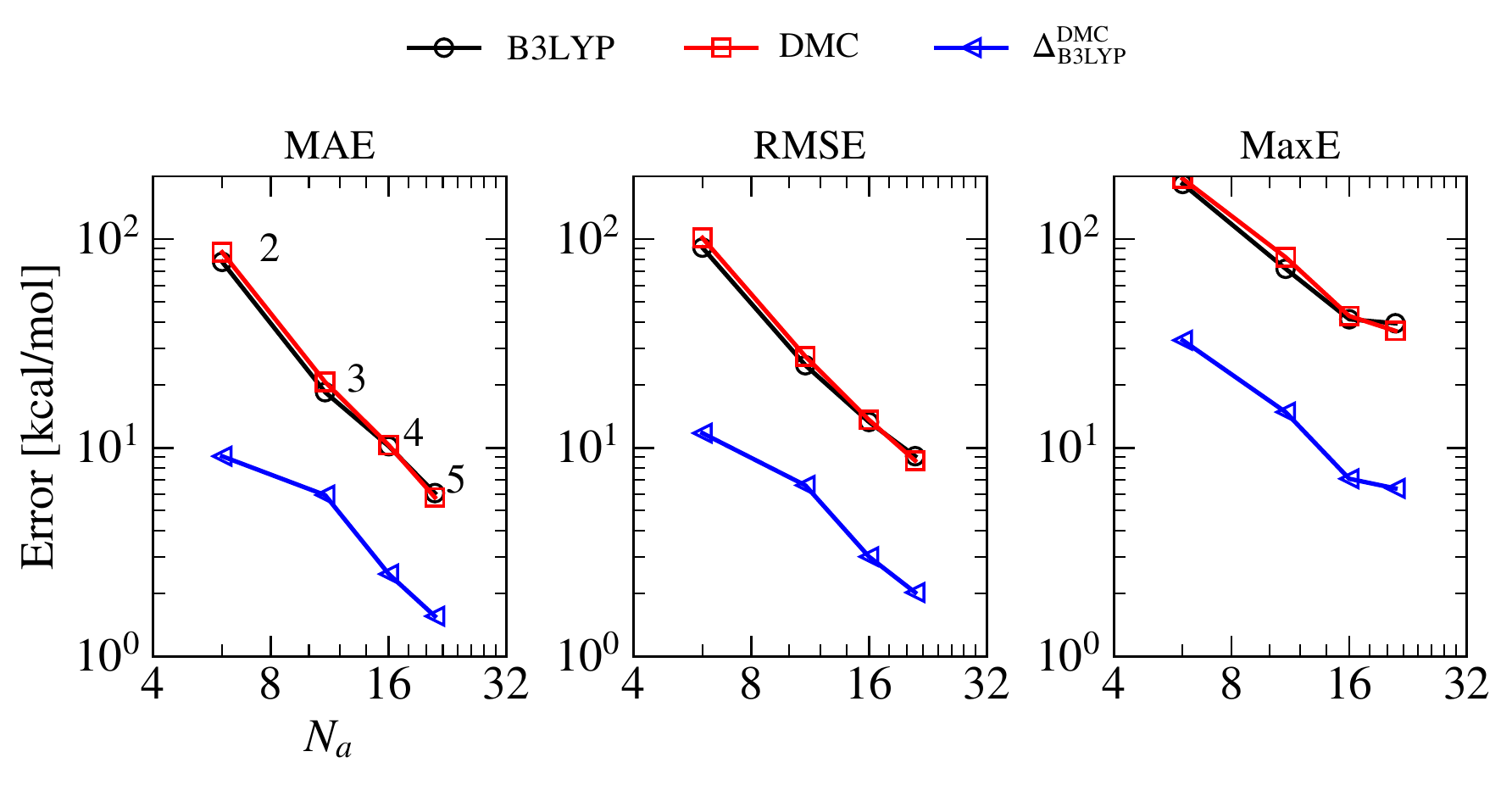}
\caption{AQML learning curves: Prediction mean absolute error (MAE), root mean squared error (RMSE) and maximal absolute error (MaxE) for estimating B3LYP or DMC total energies, or their difference ($\Delta$), of 50 test QM9 molecules (shown in Fig.~\ref{fig:test50}). 
Errors shown decay with number ($N_a$) and size ($N_I = \{2, 3, 4, 5$ heavy atoms as indicated in MAE panel) of amons used in training. 
}
\label{fig:maeLC}
\end{figure*}

The MAE's (and other error measures) of QML models at the largest training set alone is not very informative,
and more details regarding the difference in performance of AQML and $\Delta$-AQML models can only be unraveled through the analysis of their respective learning curves (LC).
Due to the similarity shared by learning curves of QML models of the same kind, we have plotted LCs for only 3 AQML models, as shown in Fig.~\ref{fig:maeLC}.

As was found in the original amons paper~\cite{Amons}, AQML models in general lead to a steepening of learning curves, much steeper than QML models based on random sampling.
A similar message is conveyed by our findings shown in Fig.~\ref{fig:maeLC}: for B3LYP trained AQML,
trained on 6 amons on average ($N_a=6$) containing no more than 2 heavy atoms ($N_I\leq 2$), the MAE is as high as $\sim$100 kcal/mol,
which quickly diminishes to $\sim$6 kcal/mol as $N_I$ grows to 21 (with $N_I\leq 5$).
Learning curves of DMC trained AQML almost overlap with that of B3LYP trained AQML (as well as the other AQML models involving correlated methods, not shown).

By adding reference energies from B3LYP as baseline to DMC trained AQML to form a $\deltabq$-AQML model along the lines of Ref.~\cite{DeltaPaper2015},
resulting learning curves shift downwards significantly. 
As already discussed in Ref.~\cite{DeltaPaper2015}, this was to be expected since the target property, deviation of DMC from B3LYP energies, is  generally lower in magnitude and in variation which makes at an easier label to learn. 
The largest drop in prediction errors (MAE, RMSE or MaxE) is observed at $N_I=2$,
with MAE reduced from $\sim$100 kcal/mol to $\sim$10 kcal/mol.
The reduction in RMSE or MaxE is roughly of the same magnitude.
This is understood as that the inclusion of amons containing 2 heavy atoms takes multivalent bonds (bond order 2 or 3) into account,
considerably improving the accuracy of the dressed-atom model
and adding more larger amons for training introduces environments that only perturb these bonds slightly.
At larger $N_I$ (or $N_a$), the slopes of both DMC trained AQML and $\deltabq$-AQML models become constant and almost the same,
since the underlying dimensionality of the learning problem has not changed~\cite{BAML,huang2020quantum}.
The constant slope also suggests that chemical accuracy will be reached soon after addition of only slightly larger amons to training. 
Unfortunately, one has to expect a roughly ten-fold increase in corresponding total amon data set size if one were to increase the number of heavy atoms in the amon training data set. 
Learning curves of RMSE and MaxE exhibit similar pattern as for MAE, the only difference being, not surprisingly, their larger magnitudes.

\subsubsection{Signed error analysis}
All analysis above provide valuable statistically averaged information about our learning problem,
but still much detail is hidden in the specific cases
and it would be advisable digging into the \emph{signed} errors for test molecule one-by-one.
It is worth pointing out that, to the best of our knowledge,
detailed (signed) error analysis is possible only for amons-based QML model, distinguishing itself from the other QML models relying on random selection of training set
(which behave, more or less, in a black box fashion).
This highly coveted feature, i.e., explainable machine learning,
endows us with the ability to understand the behavior of learning,
for instance, why the errors decrease or increase when $N_I$ ($N_a$) grows to a specific value
or why errors are large or small in magnitude, positive or negative, etc.,
and consequently we are able to know a priori how well a QML model performs for a given query molecule,
without training and test at all.
With understanding of the source of errors at this depth, 
we could systematically improve the selection scheme of amons,
as shall be elaborated.

\begin{figure}[h!]
\centering
\includegraphics[scale=0.52]{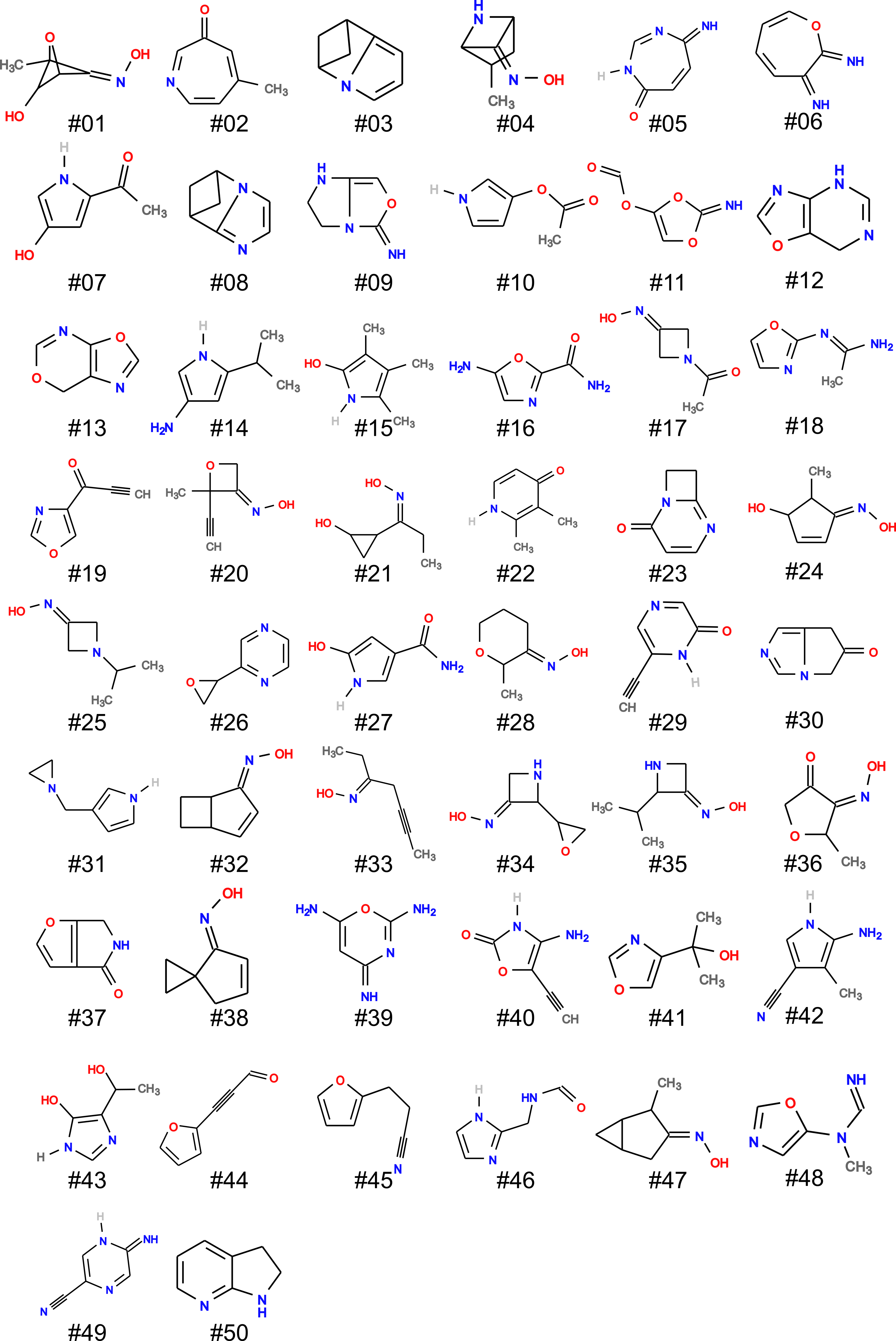} 
\caption{50 test molecules for which we performed DMC energies made up of 9 heavy atoms. Molecules were drawn at random from QM9 dataset~\cite{QM9}, and sorted by signed prediction error from largest B3LYP trained AQML model (see Fig.~\ref{fig:maeLC}), in ascending order. }
\label{fig:test50}
\end{figure}

To assist comprehensive error analysis, all test molecules (displayed in Fig.~\ref{fig:test50}) are to be divided into two main branches: strain-free (e.g., \#33) and strained.
Strain in a molecule has a multitude of origins.
Here we are concerned about only one variant of strain: the so-called angle strain (or ring strain, or Baeyer strain~\cite{wiberg1986ringstrain}),
due mostly to deviation from ideal vicinal angles.
Molecules experiencing angle strain may be further classified as small ring system (containing typically 3-5 heavy atoms, denoted as ``ST1'' for reference later),
bicyclic system (features two joined rings, denoted as ``ST2''),
or medium-sized ring system (containing 7-13 ring atoms, denoted as ``ST3'').
For instance, the QM9 molecule \#21 in Fig.~\ref{fig:test50} belongs to the ST1 class, as it contains a highly strained small cyclic structure C1CC1;
A typical ST2 molecule is \#03, in which one 5-membered ring and another 4-membered ring share two common carbon atoms, that is, joint.
Molecule \#01 consists of a 7-membered conjugate ring structure,
with angles formed by heavy atoms all over 120 degrees in the fully strain-free case, falls into the ST3 class.
A quick inspection of the 50 test QM9 molecules reveal that the majority are either strain-free, or belong to the ST1 class,
and the highest level of strain is associated with the ST2 molecules.
As we shall see, they behave rather differently within QML.
Please be noted that the categorization is not unique and some molecule can be categorized into more than one class, e.g.., \#03 belong to both ST1 and ST2.


\begin{figure*} 
\includegraphics[scale=0.56]{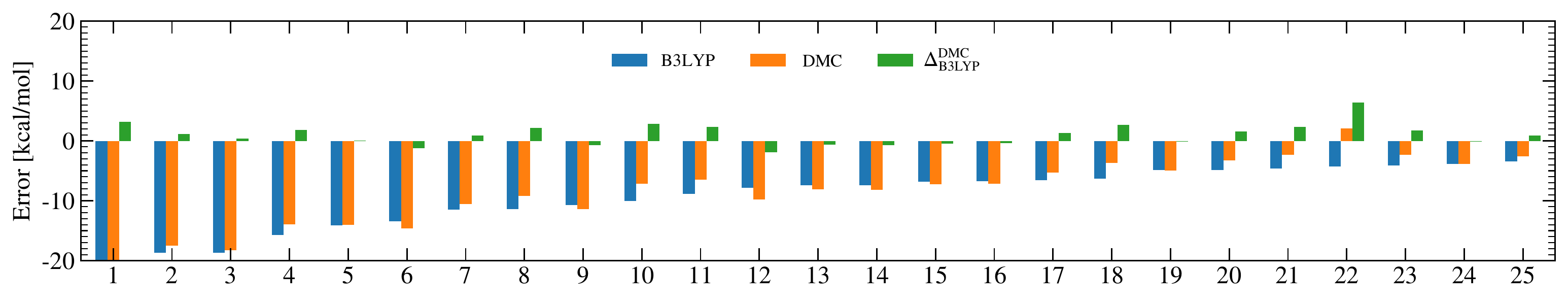}\\
\includegraphics[scale=0.56]{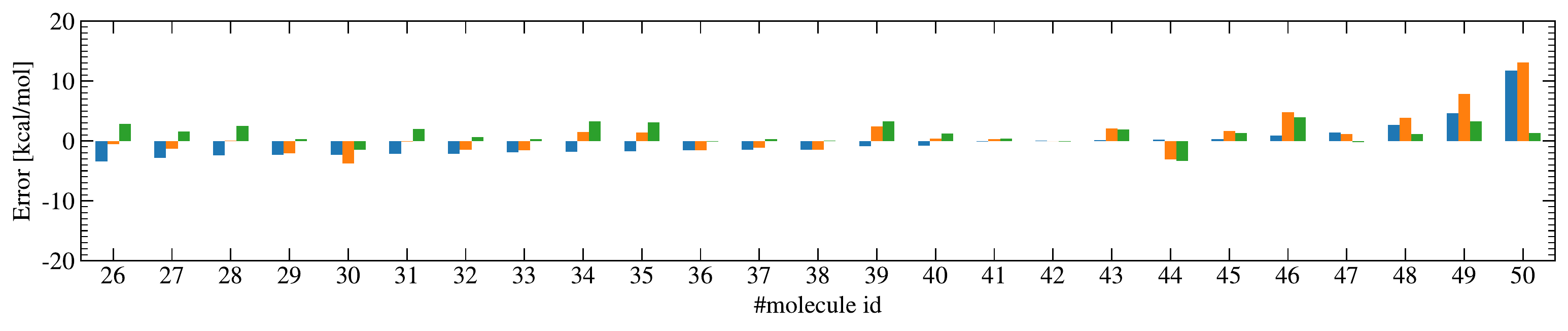}
\caption{
Bar plots of signed prediction errors for largest AQML models (see Fig.~\ref{fig:maeLC}) 
trained and predicted on B3LYP,
DMC, and $\Delta_{\rm{B3LYP}}^{\rm{DMC}}$ in blue, orange, and green, respectively.
for all 50 test molecules shown in Fig.~\ref{fig:test50}.}
\label{fig:errordistr}
\end{figure*}

Beginning with the analysis on results of the DMC trained AQML model, shown as bar plots in Fig.~\ref{fig:errordistr} for each test molecule, we note the following:
Among all test molecules, about a half of them (\#21-\#47, excluding \#46) exhibit absolute error of $\sim$ 2 kcal/mol or less.
This is encouraging as amons made up at most 5 heavy atoms are used for training and
suggests the feasibility of an amon selection scheme with adaptive $N_I$ (previously, amons with $N_I\leq 7$ are all included~\cite{Amons}) to allow for further reduction of the training set size.
As has already mentioned previously, small DMC trained AQML prediction errors
are typical for rigid molecules (i.e., no or very few rotational degrees of freedom)
that are either strain-free (e.g., \#28, \#33) or of the ST1 class (and meanwhile not belonging to either ST2 or ST3 class),
where the standalone small strained ring (usually saturated) exhibits high degrees of locality
and could be fully characterized by its amons
covering no amon conformers at all,
examples include \#31, \#34 and \#35.
If the small strained ring is coupled with other strained ring (as in the ST2 case, e.g., \#03 and \#04),
or other unsaturated bond and together the total number of heavy atoms exceeds 5
(as in \#08 and \#09),
the accuracy of the DMC trained AQML model would deteriorate without doubt.
For these outliers, the local structures in their respective amons are just too relaxed
and can hardly represent the strains experienced by the relevant atoms in the query molecules.

\begin{figure*}[th!]
\centering
\includegraphics[scale=0.8]{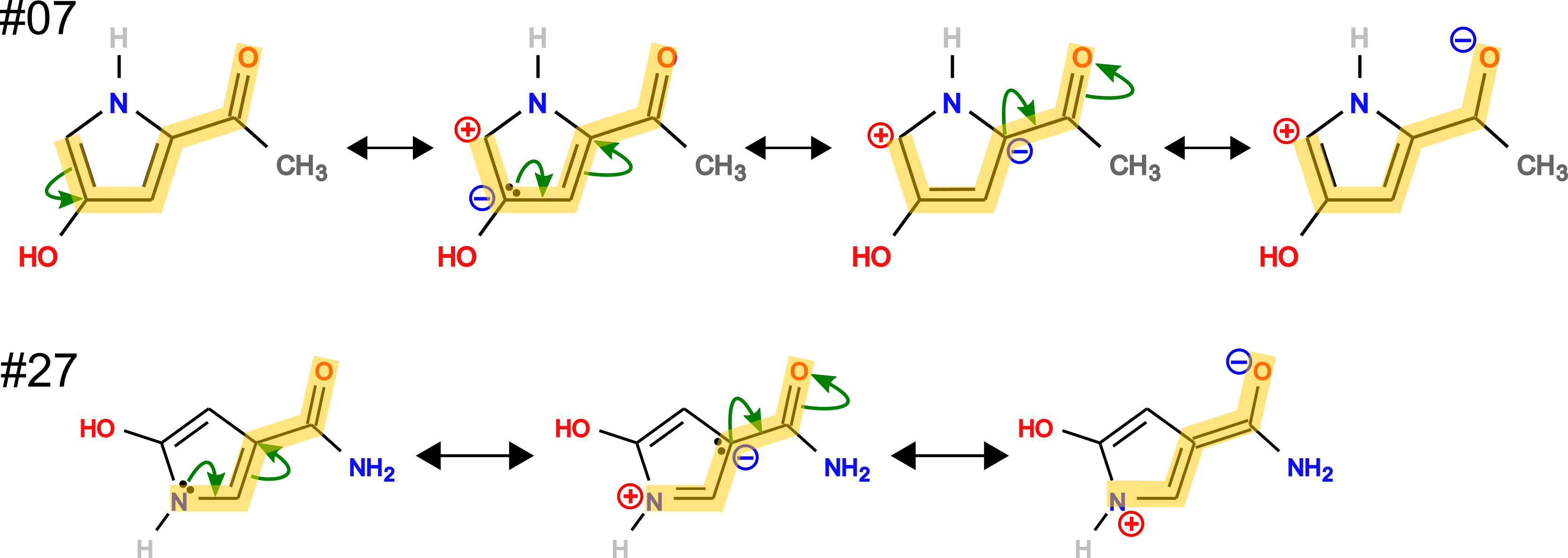}
\caption{Comparison of the resonance forms of molecules \#07 (large error of AQML model) and \#27 (small error of AQML). For the former, a maximum of 6 heavy atoms are on the resonance/conjugation path (highlighted in yellow), while 5 for the latter.}
\label{fig:resonance}
\end{figure*}

In contrast to many small error cases,
cases exhibits large error (with absolute error of more than 10 kcal/mol) are relatively much fewer,
amounting to only $\sim$1/5 of test molecules,
including 9 molecules with negative error, i.e., \#01-\#09 and one with positive error, i.e., \#50.
These molecules experience either high internal strain and/or conjugation extending over the whole molecule,
which cannot be represented by their constituting fragments (aka., amons) made up of no more 5 heavy atoms.
Of particular interest is the molecule \#07, which is similar to \#27,
but with drastic different magnitude of DMC trained AQML error, i.e., $-10.56$ kcal/mol for \#07 vs. $-1.28$ kcal/mol for \#27.
This can be explained by comparing the maximal number of heavy atoms on the conjugation path of the two molecules
and can be obtained through drawing molecular resonance forms.
As shown in Fig.~\ref{fig:resonance}, 6 atoms are on the path of the resonance forms of molecule \#07,
compared to 5 in molecule \#27.
That is, a set of amons with $\{N_I\leq 5\}$ is sufficient to describe test molecule \#27,
while $\{N_I=6\}$ amons must be present for accurate energy extrapolation of \#07.

The local structure difference in the test molecule and its amons has further striking consequence:
as shown in Fig.~\ref{fig:errordistr}, most of the AQML signed prediction errors are negative, few positive.
This naturally arises from the facts that
i) small amons by design are more relaxed (bear in mind that amon corresponds to local minima in the potential energy surface)
compared to larger test molecules, and
ii) the majority of our test molecules experience strain in one way or another.
More specifically, QML predictions for extrapolation of energies of larger molecules based on their constituent smaller amons
could be recast into essentially the interpolation between the energies of different atomic environments.
If the amons are too relaxed, the local energies of atomic environments in amons would be lower (or more negative) than that in the target molecule,
and the (weighted) summation of the interpolated atomic energies would also be too negative accordingly,
resulting in a negatively signed error,
or over stabilization seen from the training set.
Similar reasoning could also be adapted to explain the outcome of QML models for conjugation controlled systems,
for which positively signed error is expected, e.g., \#50.
That is, due to the opposite direction of change of atomic energies, 
i.e., conjugation effect (more precisely, aromaticity) introduces greater stabalization (or atomic energies are more negative) in the larger test molecule than in its amons.

In accordance with the findings in the preceding two subsections,
the errors of the B3LYP trained AQML model are similar to that of DMC trained AQML,
differing only slightly in magnitudes
for most of the test molecules,
insinuating in all likelihood the drastic reduction in prediction errors by learning the energy difference between the two levels of theory,
instead of the total energy.
This echoes exactly the spirit of the so-called $
\Delta$-ML~\cite{DeltaPaper2015}.
Not surprisingly, we have observed that
i) the error of the $\Delta$-AQML model is roughly equal to the difference between the errors of the associated AQML models;
ii) most commonly, the errors of associated AQML models are of the same sign,
subtracting one from another results in very small or negligible $\Delta$-AQML model errors;
However, there does exist few exceptions, where AQML model error signs differ.
The only noticeable example is the molecule \#22,
for which a small positive error is found for the DMC trained AQML model ($\sim$2 kcal/mol), but a (larger in absolute magnitude) negatively signed error for B3LYP trained AQML ($\sim-$4 kcal/mol),
ending up with a even greater error of $+6$ kcal/mol for the 
$\deltabq$-AQML model.
A further inspiration is that the $\Delta$-QML model would be potentially a very useful tool for identifying unusual molecules throughout the chemical compound space
for benchmarking/improving approximate quantum chemical theories, DFT in particular.

\section{Conclusion}
To summarize, we have conducted a DMC compute campaign for over 1'000 small amons containing no more than 5 heavy atoms and covering the CCS of QM9~\cite{QM9}.
Starting from this compact data-set,
we have assessed the performance of a three amons-based quantum machine learning models (AQML),
including both, conventional direct AQML models of B3LYP and DMC energies, as well as a   $\Delta$-AQML model of the difference between B3LYP and DMC.
The scalable $\deltabq$-AQML model,
utilising the atomic SLATM representation,
exhibits promising performance,
 achieving a MAE of less than 1.6 kcal/mol (DMC energy as reference) on a test set of 50 larger organic molecules drawn at random from QM9 and made up of 9 heavy atoms (not counting hydrogen).
This $\deltabq$-AQML model requires on average only $\sim$20 DMC energies for the constituting fragments/amons which are selected on the fly from the aforementioned compact data-set.
Considering that we have already achieved quite low prediction errors in this study,
and considering the fact that the learning curves indicate amenability to further reduction by either fine-tuning and extending the amon basis,
or by inclusion of even more levels of theory within multi-level grid-combination technique based QML ~\cite{zaspel2018boosting},
we believe that the amon-based $\Delta$ QML framework, in combination with costly  DMC reference datasets, hold great promise for robust yet efficient exploration campaigns of molecules and materials throughout CCS. 

Furthermore, our DMC results indicate that the nodal surface error associated to the choice of a trial wavefunction is minimal ($~$2kcal/mol) for DFT functionals. 
This finding would suggest that (at least for the sub-domain of CCS considered here within) it is justified to use DFT to generate  starting trial wavefunctions. 
Finally, we note the usefulness of relying on 
averaged dressed atomic energies, rather than proper open shell atomic references. 
This step provides a key correction towards less accurate trial wavefunctions, allowing for an unbiased evaluation of the DMC accuracy when it comes to predict molecular energies throughout CCS.


\section{Supporting information}
Geometries and HF, DFT, and DMC energies for both, the 50 test molecules (shown in Fig.~\ref{fig:test50}) as well as  the 1'175 QM9 based amons used for training. 
have been made available at the Materials Data Facility\cite{blaiszik2016materials,blaiszik2019data}, {\tt https://doi.org/10.18126/hxlp-v732 }, DOI:10.18126/hxlp-v732,
and at the Materials Cloud Archive, {\tt https://archive.materialscloud.org/}, DOI:10.24435/materialscloud:p7-p8.

\begin{acknowledgments}
O.A.v.L. has received funding from the European Research Council (ERC) under the European Union’s Horizon 2020 research and innovation programme (grant agreement No. 772834).
This research was supported by the NCCR MARVEL, a National Centre of Competence in Research, funded by the Swiss National Science Foundation (grant number 182892). O.A.v.L. acknowledge support by the Swiss National Science foundation (No.~PP00P2\_138932, 407540\_167186 NFP 75 Big Data.) \\ 

DFT and DMC calculations ran by AB and JTK who acknowledge the support of the U.S. Department of Energy, Office of Science, Basic Energy Sciences, Materials Sciences and Engineering Division, as part of the Computational Materials Sciences Program and Center for Predictive Simulation of Functional Materials.
DFT and DMC calculations used an award of computer time provided by the Innovative and Novel Computational Impact on Theory and Experiment (INCITE) program. This research has used resources of the Argonne Leadership Computing Facility, which is a DOE Office of Science User Facility supported under Contract DE-AC02-06CH11357. 

\end{acknowledgments}

\bibliographystyle{unsrt}
\bibliography{main}

\begin{thebibliography}{10}

\bibitem{huang2021abinitio}
Bing Huang and O.~Anatole von Lilienfeld.
\newblock Ab initio machine learning in chemical compound space.
\newblock {\em Chem. Rev.}, 121:10001, 2021.

\bibitem{MolecularElectronicStructureTheory}
T.~Helgaker, P.~J{\o}rgensen, and J.~Olsen.
\newblock {\em Molecular Electronic-Structure Theory}.
\newblock John Wiley \& Sons, LTD, 2000.

\bibitem{bHammond1994}
B.~Hammond, W.~Lester, and P.~Reynolds.
\newblock {\em Monte Carlo Methods in Ab Initio Quantum Chemistry}.
\newblock World Scientiﬁc, London, 1994.

\bibitem{Lester1990}
W.~A.~Jr. Lester and B.~L. Hammond.
\newblock {\em Ann. Rev. Phys. Chem.(Palo Alto, CA: Annual Reviews)}, 41:283,
  1990.

\bibitem{foulkes01}
W.~M.~C Foulkes, L.~Mitas, R.~J. Needs, and G.~Rajagopal.
\newblock Quantum monte carlo simulations of solids.
\newblock {\em Rev.\ Mod.\ Phys.}, 73:33--83, 2001.

\bibitem{Metropolis1953}
Nicholas Metropolis, Arianna~W. Rosenbluth, Marshall~N. Rosenbluth, Augusta~H.
  Teller, and Edward Teller.
\newblock Equation of state calculations by fast computing machines.
\newblock {\em The Journal of Chemical Physics}, 21(6):1087--1092, 1953.

\bibitem{Reynolds1982}
P.~J. Reynolds, D.~M. Ceperley, B.~J. Alder, and W.~A. Lester.
\newblock Fixed-node quantum monte carlo for molecules.
\newblock {\em J. Chem. Phys.}, 77:5593--5603, Jun 1982.

\bibitem{Anderson1980}
James~B. Anderson.
\newblock Quantum chemistry by random walk: Higher accuracy.
\newblock {\em The Journal of Chemical Physics}, 73(8):3897--3899, 1980.

\bibitem{Ceperley1980}
D.~M. Ceperley and B.~J. Alder.
\newblock Ground state of the electron gas by a stochastic method.
\newblock {\em Phys. Rev. Lett.}, 45(7):566--569, Aug 1980.

\bibitem{Wagner2017}
Kittithat Krongchon, Brian Busemeyer, and Lucas~K. Wagner.
\newblock Accurate barrier heights using diffusion monte carlo.
\newblock {\em The Journal of Chemical Physics}, 146(12):124129, 2017.

\bibitem{Dubecky2017}
Mat\'u\ifmmode \check{s}\else~\v{s}\fi{} Dubeck\'y.
\newblock Bias cancellation in one-determinant fixed-node diffusion monte
  carlo: Insights from fermionic occupation numbers.
\newblock {\em Phys. Rev. E}, 95:033308, Mar 2017.

\bibitem{Burke2018}
Suhwan Song, Min-Cheol Kim, Eunji Sim, Anouar Benali, Olle Heinonen, and Kieron
  Burke.
\newblock Benchmarks and reliable dft results for spin gaps of small ligand
  fe(ii) complexes.
\newblock {\em Journal of Chemical Theory and Computation}, 14(5):2304--2311,
  2018.
\newblock PMID: 29614856.

\bibitem{Dubecky2013}
Matúš Dubecký, Petr Jurečka, René Derian, Pavel Hobza, Michal Otyepka, and
  Lubos Mitas.
\newblock Quantum monte carlo methods describe noncovalent interactions with
  subchemical accuracy.
\newblock {\em Journal of Chemical Theory and Computation}, 9(10):4287--4292,
  2013.
\newblock PMID: 26589147.

\bibitem{shulenburger2015}
L.~Shulenburger, A.D. Baczewski, Z.~Zhu, J.~Guan, and D.~Tománek.
\newblock The nature of the interlayer interaction in bulk and few-layer
  phosphorus.
\newblock {\em Nano Letters}, 15(12):8170--8175, 2015.
\newblock PMID: 26523860.

\bibitem{sorella2015}
N.~Devaux, M.~Casula, F.~Decremps, and S.~Sorella.
\newblock Electronic origin of the volume collapse in cerium.
\newblock {\em Phys. Rev. B}, 91:081101, Feb 2015.

\bibitem{Foulkes2001}
W.~M.~C. Foulkes, L.~Mitas, R.~J. Needs, and G.~Rajagopal.
\newblock Quantum monte carlo simulations of solids.
\newblock {\em Rev. Mod. Phys.}, 73:33--83, Jan 2001.

\bibitem{Shulenburger2013a}
Luke Shulenburger and Thomas~R. Mattsson.
\newblock Quantum monte carlo applied to solids.
\newblock {\em Phys. Rev. B}, 88:245117, Dec 2013.

\bibitem{ArQMC_AnouarJCTC2014}
A.~Benali, N.~A. Romero, L.~Shulenburger, J.~Kim, and O.~A. von Lilienfeld.
\newblock {Application of diffusion Monte Carlo to materials dominated by van
  der Waals interactions}.
\newblock {\em {J. Chem. Theory Comput.}}, 10:3417, 2014.

\bibitem{Morales2012}
Miguel~A. Morales, Jeremy McMinis, Bryan~K. Clark, Jeongnim Kim, and Gustavo~E.
  Scuseria.
\newblock Multideterminant wave functions in quantum monte carlo.
\newblock {\em Journal of Chemical Theory and Computation}, 8(7):2181--2188,
  2012.

\bibitem{caffarel2016}
Michel Caffarel, Thomas Applencourt, Emmanuel Giner, and Anthony Scemama.
\newblock Communication: Toward an improved control of the fixed-node error in
  quantum monte carlo: The case of the water molecule.
\newblock {\em The Journal of Chemical Physics}, 144(15):151103, 2016.

\bibitem{Scemama2018}
Anthony Scemama, Anouar Benali, Denis Jacquemin, Michel Caffarel, and
  Pierre-François Loos.
\newblock Excitation energies from diffusion monte carlo using selected
  configuration interaction nodes.
\newblock {\em The Journal of Chemical Physics}, 149(3):034108, 2018.

\bibitem{Wagner2013}
Lucas~K. Wagner.
\newblock Types of single particle symmetry breaking in transition metal oxides
  due to electron correlation.
\newblock {\em The Journal of Chemical Physics}, 138(9):094106, 2013.

\bibitem{Amons}
Bing Huang and O.~Anatole von Lilienfeld.
\newblock Quantum machine learning using atom-in-molecule-based fragments
  selected on the fly.
\newblock {\em Nature Chemistry}, 12(10):945--951, October 2020.

\bibitem{QM9}
R.~Ramakrishnan, P.~Dral, M.~Rupp, and O.~A. von Lilienfeld.
\newblock Quantum chemistry structures and properties of 134 kilo molecules.
\newblock {\em Scientific Data}, 1:140022, 2014.

\bibitem{ReymondChemicalUniverse3}
L.~C. Blum and J.-L. Reymond.
\newblock 970 million druglike small molecules for virtual screening in the
  chemical universe database {GDB-13}.
\newblock {\em J. Am. Chem. Soc.}, 131:8732, 2009.

\bibitem{rdkit}
RDKit: Open-source cheminformatics; http://www.rdkit.org.

\bibitem{g09}
M.~J. Frisch, G.~W. Trucks, H.~B. Schlegel, G.~E. Scuseria, M.~A. Robb, J.~R.
  Cheeseman, G.~Scalmani, V.~Barone, B.~Mennucci, G.~A. Petersson,
  H.~Nakatsuji, M.~Caricato, X.~Li, H.~P. Hratchian, A.~F. Izmaylov, J.~Bloino,
  G.~Zheng, J.~L. Sonnenberg, M.~Hada, M.~Ehara, K.~Toyota, R.~Fukuda,
  J.~Hasegawa, M.~Ishida, T.~Nakajima, Y.~Honda, O.~Kitao, H.~Nakai, T.~Vreven,
  J.~A. Montgomery, Jr., J.~E. Peralta, F.~Ogliaro, M.~Bearpark, J.~J. Heyd,
  E.~Brothers, K.~N. Kudin, V.~N. Staroverov, R.~Kobayashi, J.~Normand,
  K.~Raghavachari, A.~Rendell, J.~C. Burant, S.~S. Iyengar, J.~Tomasi,
  M.~Cossi, N.~Rega, J.~M. Millam, M.~Klene, J.~E. Knox, J.~B. Cross,
  V.~Bakken, C.~Adamo, J.~Jaramillo, R.~Gomperts, R.~E. Stratmann, O.~Yazyev,
  A.~J. Austin, R.~Cammi, C.~Pomelli, J.~W. Ochterski, R.~L. Martin,
  K.~Morokuma, V.~G. Zakrzewski, G.~A. Voth, P.~Salvador, J.~J. Dannenberg,
  S.~Dapprich, A.~D. Daniels, \"O. Farkas, J.~B. Foresman, J.~V. Ortiz,
  J.~Cioslowski, and D.~J. Fox.
\newblock Gaussian˜09, 2009.
\newblock Gaussian Inc. Wallingford CT.

\bibitem{molpro}
H.-J. Werner, P.~J. Knowles, G.~Knizia, F.~R. Manby, M.~{Sch\"{u}tz},
  P.~Celani, W.~Gy\"orffy, D.~Kats, T.~Korona, R.~Lindh, A.~Mitrushenkov,
  G.~Rauhut, K.~R. Shamasundar, T.~B. Adler, R.~D. Amos, S.~J. Bennie,
  A.~Bernhardsson, A.~Berning, D.~L. Cooper, M.~J.~O. Deegan, A.~J. Dobbyn,
  F.~Eckert, E.~Goll, C.~Hampel, A.~Hesselmann, G.~Hetzer, T.~Hrenar,
  G.~Jansen, C.~K\"oppl, S.~J.~R. Lee, Y.~Liu, A.~W. Lloyd, Q.~Ma, R.~A. Mata,
  A.~J. May, S.~J. McNicholas, W.~Meyer, T.~F. {Miller III}, M.~E. Mura,
  A.~Nicklass, D.~P. O'Neill, P.~Palmieri, D.~Peng, K.~Pfl\"uger, R.~Pitzer,
  M.~Reiher, T.~Shiozaki, H.~Stoll, A.~J. Stone, R.~Tarroni, T.~Thorsteinsson,
  M.~Wang, and M.~Welborn.
\newblock Molpro, version 2019.2, a package of ab initio programs, 2019.
\newblock https://www.molpro.net.

\bibitem{Schmidt1990}
K.~E. Schmidt and J.~W. Moskowitz.
\newblock Correlated monte carlo wave functions for the atoms he through ne.
\newblock {\em The Journal of Chemical Physics}, 93(6):4172--4178, 1990.

\bibitem{kim2018qmcpack}
Jeongnim Kim, Andrew~T Baczewski, Todd~D Beaudet, Anouar Benali, M~Chandler
  Bennett, Mark~A Berrill, Nick~S Blunt, Edgar Josu{\'e}~Landinez Borda,
  Michele Casula, David~M Ceperley, et~al.
\newblock Qmcpack: an open source ab initio quantum monte carlo package for the
  electronic structure of atoms, molecules and solids.
\newblock {\em Journal of Physics: Condensed Matter}, 30(19):195901, 2018.

\bibitem{kent2020qmcpack}
P.~R.~C. Kent, Abdulgani Annaberdiyev, Anouar Benali, M.~Chandler Bennett,
  Edgar~Josué Landinez~Borda, Peter Doak, Hongxia Hao, Kenneth~D. Jordan,
  Jaron~T. Krogel, Ilkka Kylänpää, Joonho Lee, Ye~Luo, Fionn~D. Malone,
  Cody~A. Melton, Lubos Mitas, Miguel~A. Morales, Eric Neuscamman, Fernando~A.
  Reboredo, Brenda Rubenstein, Kayahan Saritas, Shiv Upadhyay, Guangming Wang,
  Shuai Zhang, and Luning Zhao.
\newblock Qmcpack: Advances in the development, efficiency, and application of
  auxiliary field and real-space variational and diffusion quantum monte carlo.
\newblock {\em The Journal of Chemical Physics}, 152(17):174105, 2020.

\bibitem{PBE}
J.~P. Perdew, K.~Burke, and M.~Ernzerhof.
\newblock Generalized gradient approximation made simple.
\newblock {\em Phys. Rev. Lett.}, 77:3865, 1996.

\bibitem{PBE0}
J.~P. Perdew, M.~Ernzerhof, and K.~Burke.
\newblock {\em J. Chem. Phys.}, 105:9982, 1996.

\bibitem{B3LYP1}
Axel~D. Becke.
\newblock Density‐functional thermochemistry. iii. the role of exact
  exchange.
\newblock {\em The Journal of Chemical Physics}, 98(7):5648--5652, 1993.

\bibitem{B3LYP2}
Chengteh Lee, Weitao Yang, and Robert~G. Parr.
\newblock Development of the colle-salvetti correlation-energy formula into a
  functional of the electron density.
\newblock {\em Phys. Rev. B}, 37:785--789, Jan 1988.

\bibitem{B3LYP3}
S.~H. Vosko, L.~Wilk, and M.~Nusair.
\newblock Accurate spin-dependent electron liquid correlation energies for
  local spin density calculations: a critical analysis.
\newblock {\em Canadian Journal of Physics}, 58(8):1200--1211, 1980.

\bibitem{B3LYP4}
P.~J. Stephens, F.~J. Devlin, C.~F. Chabalowski, and M.~J. Frisch.
\newblock Ab initio calculation of vibrational absorption and circular
  dichroism spectra using density functional force fields.
\newblock {\em The Journal of Physical Chemistry}, 98(45):11623--11627, 1994.

\bibitem{pyscf}
Qiming Sun, Timothy~C. Berkelbach, Nick~S. Blunt, George~H. Booth, Sheng Guo,
  Zhendong Li, Junzi Liu, James~D. McClain, Elvira~R. Sayfutyarova, Sandeep
  Sharma, Sebastian Wouters, and Garnet~Kin‐Lic Chan.
\newblock Pyscf: the python‐based simulations of chemistry framework, 2017.

\bibitem{umrigar07}
C.~J. Umrigar, Julien Toulouse, Claudia Filippi, S.~Sorella, and R.~G. Hennig.
\newblock Alleviation of the fermion-sign problem by optimization of many-body
  wave functions.
\newblock {\em Phys. Rev. Lett.}, 98:110201, Mar 2007.

\bibitem{DeltaPaper2015}
R.~Ramakrishnan, P.~Dral, M.~Rupp, and O.~A. von Lilienfeld.
\newblock {Big Data meets Quantum Chemistry Approximations: The
  $\Delta$-Machine Learning Approach}.
\newblock {\em J. Chem. Theory Comput.}, 11:2087, 2015.

\bibitem{pfau2019abinitio}
David Pfau, James~S. Spencer, Alexander~G. de~G.~Matthews, and W.~M.~C.
  Foulkes.
\newblock Ab-initio solution of the many-electron schrödinger equation with
  deep neural networks, 2019.

\bibitem{Busemeyer2016}
Brian Busemeyer, Mario Dagrada, Sandro Sorella, Michele Casula, and Lucas~K.
  Wagner.
\newblock Competing collinear magnetic structures in superconducting fese by
  first-principles quantum monte carlo calculations.
\newblock {\em Phys. Rev. B}, 94:035108, Jul 2016.

\bibitem{thygesen2005partly}
Kristian~S Thygesen, Lars~Bruno Hansen, and Karsten~Wedel Jacobsen.
\newblock Partly occupied wannier functions.
\newblock {\em Physical review letters}, 94(2):026405, 2005.

\bibitem{BAML}
Bing Huang and O.~Anatole von Lilienfeld.
\newblock Communication: Understanding molecular representations in machine
  learning: The role of uniqueness and target similarity.
\newblock {\em J. Chem. Phys.}, 145(16), 2016.

\bibitem{huang2020quantum}
Bing Huang, Nadine~O Symonds, and O~Anatole von Lilienfeld.
\newblock Quantum machine learning in chemistry and materials.
\newblock {\em Handbook of Materials Modeling: Methods: Theory and Modeling},
  pages 1883--1909, 2020.

\bibitem{wiberg1986ringstrain}
Kenneth~B Wiberg.
\newblock The concept of strain in organic chemistry.
\newblock {\em Angewandte Chemie International Edition in English},
  25(4):312--322, 1986.

\bibitem{zaspel2018boosting}
Peter Zaspel, Bing Huang, Helmut Harbrecht, and O~Anatole von Lilienfeld.
\newblock Boosting quantum machine learning models with multi-level combination
  technique: Pople diagrams revisited.
\newblock {\em Journal of chemical theory and computation}, 2018.

\bibitem{blaiszik2016materials}
Ben Blaiszik, Kyle Chard, Jim Pruyne, Rachana Ananthakrishnan, Steven Tuecke,
  and Ian Foster.
\newblock The materials data facility: data services to advance materials
  science research.
\newblock {\em Jom}, 68(8):2045--2052, 2016.

\bibitem{blaiszik2019data}
Ben Blaiszik, Logan Ward, Marcus Schwarting, Jonathon Gaff, Ryan Chard, Daniel
  Pike, Kyle Chard, and Ian Foster.
\newblock A data ecosystem to support machine learning in materials science.
\newblock {\em MRS Communications}, 9(4):1125--1133, 2019.

\end{thebibliography}
\end{document}